\documentclass{aa}  
\usepackage{psfig}
\usepackage{amssymb}
\usepackage{graphicx}
\usepackage[authoryear]{natbib}
\usepackage[figuresright]{rotating}
\usepackage{txfonts}

\newcommand{\teff}{\ifmmode T_{\rm eff} \else $T_{\mathrm{eff}}$\fi}
\newcommand{\logg}{\ifmmode \log g \else $\log g$\fi}
\newcommand{\msun}{\ifmmode M_{\odot} \else M$_{\odot}$\fi}
\newcommand{\rsun}{\ifmmode R_{\odot} \else R$_{\odot}$\fi}
\newcommand{\zsun}{\ifmmode Z_{\odot} \else Z$_{\odot}$\fi}
\newcommand{\lsun}{\ifmmode L_{\odot} \else L$_{\odot}$\fi}

\newcommand{\mdot}{\ifmmode \dot{M} \else $\dot{M}$\fi}
\newcommand{\vinf}{\ifmmode v_{\infty} \else $v_{\infty}$\fi}
\newcommand{\vturb}{\ifmmode v_{\rm turb} \else $v_{\rm turb}$\fi}
\newcommand{\mv}{\ifmmode M_{\rm{V}} \else M$_{\rm{V}}$\fi}
\newcommand{\lL}{\ifmmode \log \frac{L}{L_{\odot}} \else $\log \frac{L}{L_{\odot}}$\fi}
\newcommand{\kms}{km s$^{-1}$}
\newcommand{\myr}{M$_{\odot}$ yr$^{-1}$}

\def\aap{A\&A}
\def\aaps{A\&AS}

\def\apj{ApJ}
\def\apjl{ApJ}
\def\apjs{ApJS}
\def\mnras{MNRAS}

\begin{document}
   \title{UBVJHK synthetic photometry of Galactic O stars}

   \subtitle{}

   \author{Fabrice Martins\inst{1}
          \and
          Bertrand Plez\inst{2}
          }

   \offprints{F. Martins}

   \institute{Max-Planck Institut f$\ddot{\rm u}$r Extraterrestrische Physik, Postfach-1312, D-85741, Garching, Germany \\
              \email{martins@mpe.mpg.de}
         \and
             GRAAL, CNRS UMR 5024, Universit\'e Montpellier II, F-34095 Montpellier Cedex 5, France\\
             \email{bertrand.plez@graal.univ-montp2.fr}
             }

   \date{Received 2 June 2006 / Accepted 16 June 2006}

\titlerunning{Synthetic photometry of O stars}
\authorrunning{F. Martins \& B. Plez}

 
  \abstract
   {} 
   {The development of powerful infrared observational technics enables the study of very extincted objects and young embedded star forming regions. This is especially interesting in the context of massive stars which form and spend a non negligible fraction of their life still enshrouded in their parental molecular cloud. Spectrophotometric calibrations are thus necessary to constrain the physical properties of heavily extincted objects.}
   {Here, we derive UBVJHK magnitudes and bolometric corrections from a grid of atmosphere models for O stars. Bessel passbands are used. Bolometric corrections ($BC$) are derived as a function of \teff\ and are subsequently used to derive $BC -$ spectral type ($ST$) and Absolute Magnitudes $- ST$ relations.}
   {Infrared magnitudes and, for the first time, bolometric corrections are given for the full range of spectral types and luminosity classes. Infrared colors are essentially constant. $(H-K)_{0}$ is 0.05 mag bluer than previously proposed. Optical calibrations are also provided and are similar to previous work, except for $(B-V)_{0}$ which is found to be at minimum -0.28 for standard O stars, slightly larger (0.04 mag) than commonly accepted.}
   {We present a consistent set of photometric calibrations of optical and infrared magnitudes and bolometric corrections for Galactic O stars as a function of \teff\ and spectral type based on non-LTE atmosphere models including winds and line-blanketing.}


   \maketitle

%
%
\section{Introduction}
\label{intro}

Massive stars are known to play an important role in various fields of
astrophysics, from stellar physics to ISM studies and chemical
evolution of galaxies, and to cosmological issues such as the
reionisation of the Universe. In particular, the connection between
massive stars and star formation is very tight: as a result of their
short lifetimes, massive stars are associated with star forming events,
and their feedback effects (radiation, winds) have a strong impact on
star formation processes. Moreover, their ionising fluxes are
responsible for nebular emission lines such as Ly$_{\alpha}$ or
H$_{\alpha}$, two lines usually used to trace star formation
\citep{kennicutt,russeil}. However, the details of the formation of
\textit{individual} massive stars is still a matter of debate: a
standard accretion process faces the problem of the strong radiative
pressure generated by the luminosity of young massive proto-stars, so
that the mass growth can be stopped at around 10 \msun. Although
progress has been recently made \citep{ys02,krumholz}, another
scenario in which massive stars form through mergers of low mass
protostars in dense clusters was proposed by \citet{bbz98}. This key
question of the formation of the most massive stars has triggered a
number of observational studies aimed at obtaining constraints on the
properties of the youngest objects \citep[e.g.][]{pauluchii,bik}. Due
to the short evolutionary timescale of massive stars, heavily
extincted young star forming regions have to be probed, which requires
the use of infrared spectrophotometry.

Although in principle using only spectroscopy allows a derivation of
spectral types and luminosity classes (LC), photometry can be
useful. This is the case when spectra have to be corrected for nebular
emission always present in star forming regions, rendering the line
strength/shape uncertain. As a result, spectral classification and
luminosity classes determinations are difficult. Moreover the
luminosity is usually derived from observed magnitudes,
extinction and bolometric corrections. Estimates of extinction often
rely on intrinsic colors of stars while the knowledge of bolometric
corrections requires atmosphere models. Hence, accurate intrinsic
photometry is crucial to get access to luminosities. Although such photometry
is usually available in the optical range \citep{kur79,sk82,cgm86},
this is not the case in the infrared where calibrations are
incomplete. The widely used intrinsic colors of \citet{kor83} are
only given for O6 to O9.5 dwarfs and the latest O
supergiants. \citet{johnson66} covers the same range of spectral types
/ luminosity class.

In this context, the recent development of reliable atmosphere models
for massive stars is certainly welcome. Indeed, the inclusion of
line-blanketing in such models now allows realistic prediction of
atmospheric structures and emergent spectra which are used to get
quantitative constraints on the properties of massive stars
\citep{paul02,hil03,jc03,martins04,repolust04,martins05}. The grid of
models computed by \citet{msh05} (herafter MSH05) and the associated
SEDs are especially interesting since they can be used to compute
optical and, most importantly, near infrared photometry for the whole
range of O stars. Together with effective temperature scales,
calibrations of magnitudes and bolometric corrections as a function of
spectral type can thus be produced.

In this paper, we have used the SEDs of MSH05 to calculate UBVJHK
photometry. In Sect.\ \ref{synth_phot} we
present our method and gives the results which are discussed in Sect.\
\ref{comp_previous} and summarised in Sect.\ \ref{conclusion}.

%
%
\section{Synthetic photometry}
\label{synth_phot}

\subsection{Method and results}
\label{meth_res}

Synthetic photometry has been computed from the grid of atmosphere
models presented by \citet{msh05} \footnote{Note that the spectral
energy distributions of these models are available at the following
URL http://www.mpe.mpg.de/$\sim$martins/SED.html}. From the emergent
SED (flux per unit of star surface, $F_{\lambda}$) we computed the
magnitude in each band

\begin{equation}
\label{eq_mag}
M_{\lambda} = -2.5 \times \log \int{F_{\lambda} B_{\lambda} d\lambda} + constant
\end{equation}

\noindent with $B_{\lambda}$ the filter passband, according to
\citet{bcp98}. Photometry is thus computed in the UBV passbands
defined by \citet{bessel90} (Johnson-Cousins system) and in the JHK
passbands of \citet{bb88} (Johnson-Glass system) for the near-IR.

\begin{figure}[t]
\centerline{\psfig{file=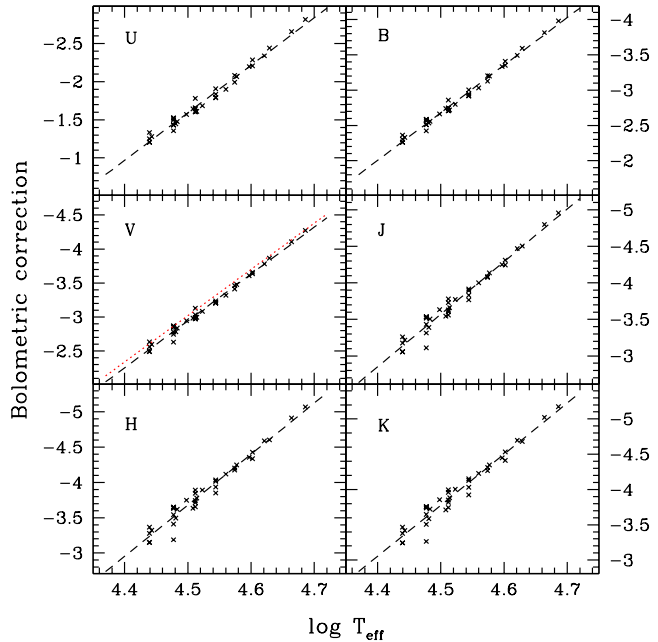,width=9cm}}
\caption{Bolometric correction in different bands as a function of
effective temperature for the models of MSH05 (crosses). The dashed
lines are the regression curves (see Table \ref{bc_teff_param} for the
corresponding parameters). The red dotted line in the plot $BC_{\rm V}
-$ \teff\ plot is the relation given by MSH05.  }
\label{BC_teff}
\end{figure}

Once obtained, these magnitudes were subsequently used to determine
bolometric corrections for each band from 

\begin{equation}
\label{eq_bc_mag}
BC_{\rm \lambda} = M_{\odot}^{bol}-M_{\lambda}-2.5 \times
\log(\frac{L}{L_{\odot}})
\end{equation}

\noindent where $ BC_{\rm \lambda}$ is the associated bolometric
correction, $M_{\odot}^{bol}$ is the bolometric magnitude of the sun
taken to be equal to 4.75 \citep[recommendation of][]{iau99} and $L$
is the luminosity.  $BC_{\rm \lambda}$ is thus computed for each model
so that we have bolometric corrections for the whole range of
effective temperatures of O stars. These values are shown in Fig.\
\ref{BC_teff}: we see that, for each band, there is a tight
correlation between $BC_{\rm \lambda}$ and $\log \teff$. A simple linear
regression of the form

\begin{equation}
\label{eq_reg_bc}
BC_{\rm \lambda} = A \times \log(\teff) + B
\end{equation}

\noindent gives the calibration between $BC_{\rm \lambda}$ and effective
temperature. Parameters $A$ and $B$ of Eq.\ \ref{eq_reg_bc} are given in
Table \ref{bc_teff_param} together with the dispersion $\sigma$. The
dispersion is larger in the infrared due to the increasing
contribution of wind emission which introduces a scatter (see Sect.\
\ref{uncertainties_cal}).

\begin{figure*}
\begin{minipage}[c]{1.0\textwidth}
\centerline{\hbox{
                  \psfig{file=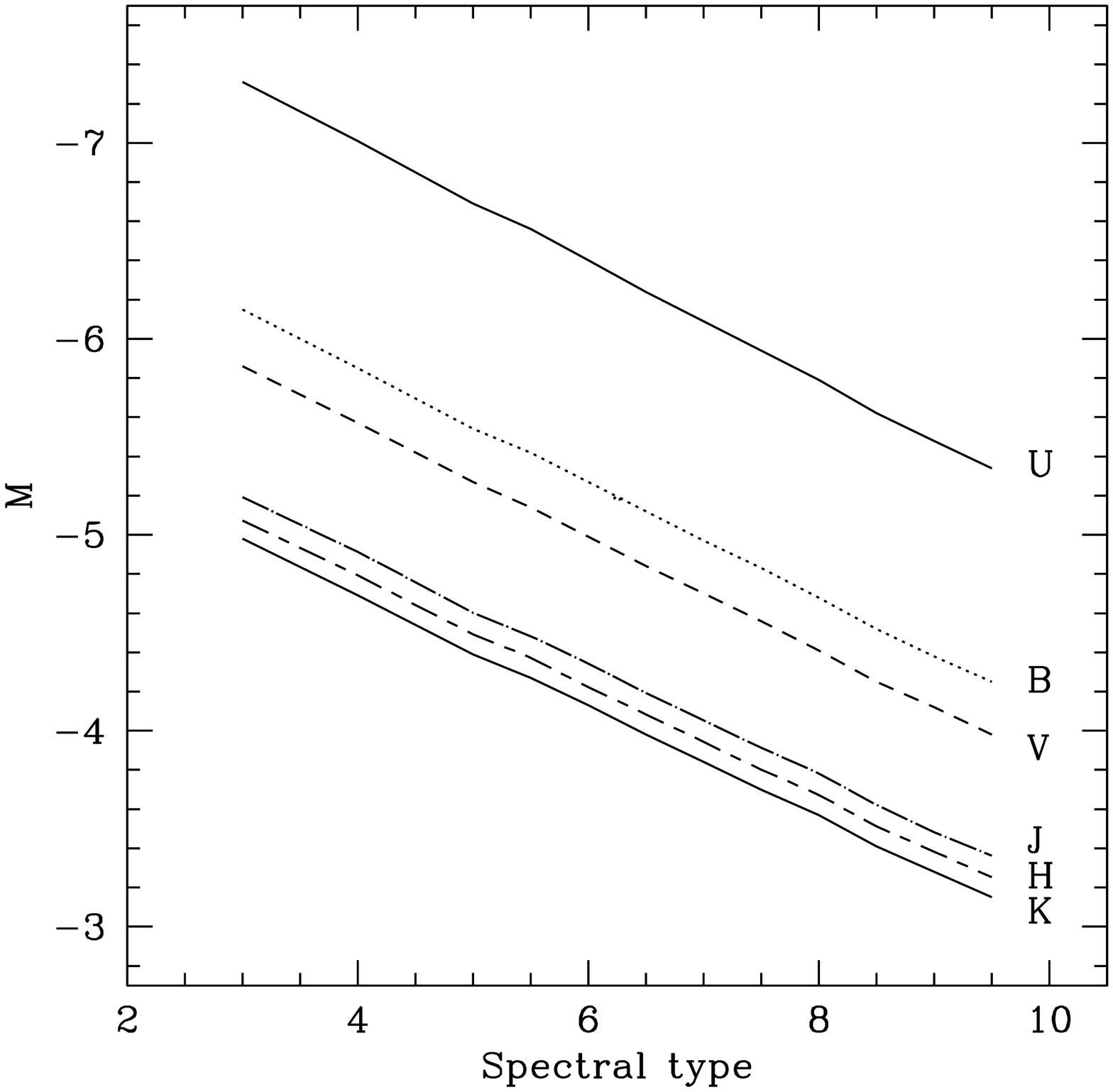,width=9.0cm}
                  \hspace{0.0cm}
                  \psfig{file=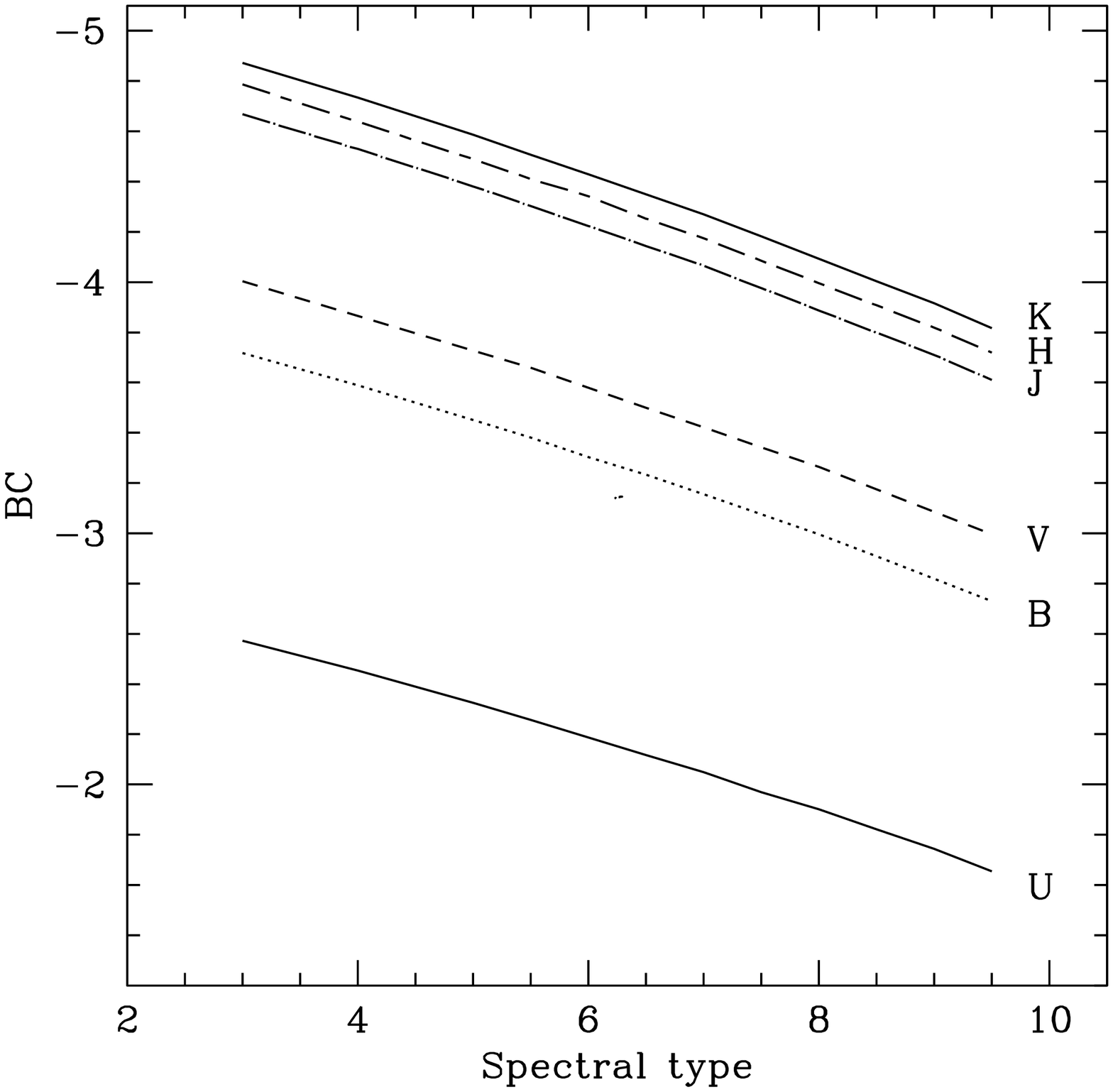,width=9.0cm}
                 }}
\vspace{0.0cm}
\hbox{\hspace{3.5cm} (a) \hspace{8.0cm} (b)} 
\vspace{0.1cm}
\caption{Magnitudes (left) and bolometric corrections (right) as a function of spectral type for dwarf stars computed using the ``observational'' \teff\ - scale of MSH05. 
}
\label{mbc_st_v}
\end{minipage}
\end{figure*}

\begin{table}
\caption{Parameters of the linear regression curves for the $BC_{\rm
\lambda}$ - \teff\ relations ($A$, $B$, see Eq.\ \ref{eq_reg_bc})
together with dispersion $\sigma$.}
\label{bc_teff_param}
\center
\begin{tabular}{l|lll}
\hline 
Band & $A$ & $B$ & $\sigma$ \\
\hline
U & -6.23 & 26.46 & 0.05 \\
B & -6.75 & 27.70 & 0.04 \\
V & -6.89 & 28.07 & 0.05 \\
J & -7.20 & 28.85 & 0.08 \\
H & -7.24 & 28.89 & 0.09 \\
K & -7.24 & 28.80 & 0.10 \\
\hline 
\end{tabular}
\end{table}

The \teff\ - scales of MSH05 for various luminosity
classes were subsequently used to convert effective temperatures into
spectral types, leading to the calibration of bolometric corrections
as a function of spectral type. The results of this simple conversion
using the ``observational'' \teff\ - scale of MSH05 are shown in Fig.\
\ref{mbc_st_v} for dwarf stars.

Finally, using these new $BC_{\rm \lambda} - ST$ calibrations, the
relations between $\log \frac{L}{L_{\odot}}$ and $ST$ of MSH05 and
Eq.\ \ref{eq_bc_mag}, we can estimate the absolute magnitude in all
bands for each spectral type and luminosity classes: we obtain the
$M_{\rm \lambda} - ST$ relations (Fig.\ \ref{mbc_st_v} shows the relation
for dwarf stars).

To avoid any confusion, we want to stress that using Eq.\ \ref{eq_mag}
gives directly the absolute magnitudes for each model, and in
principle inspection of the spectrum can also provide the spectral
type. In practice, we then have $M_{\rm \lambda}$ and the spectral
type for each model. However, we are here interested in producing
calibrations $BC_{\rm \lambda} - ST$ and $M_{\rm \lambda} - ST$ for
the three main luminosity classes (dwarfs, giants and supergiants)
which are defined by specific relations between $\logg$ and $ST$ (see
MSH05). Our models do not fall exactly on these relations, some being
in between two luminosity classes (by construction of the
grid). Hence, the approach adopted here is aimed at taking this into
account and is well suited for our purpose, namely producing
calibrations for each luminosity class.

The results of the calibrations are gathered in Tables \ref{tab_obs}
and \ref{tab_theo}. In the former, the ``observational'' \teff\ scale
of MSH05 is used, while in the latter, we make use of their
``theoretical'' \teff\ - ST relation. Inspection of Tables
\ref{tab_obs} and \ref{tab_theo} reveals that adopting the
``theoretical'' \teff\ scale of MSH05 changes very little the
magnitudes: the differences are not larger than 0.03 mag. The
conclusion is the same for bolometric corrections of the earliest
spectral types, while later type stars suffer from larger differences
(up to 0.22 mag). This is expected since late types are the ones for
which the two effective temperature scales differ the most. Note that
the colors are basically unaffected by the change of effective
temperature scale since they depend very little on \teff. Since at
present it is not clear which \teff\ - scale better reflects the true
properties of O stars, we chose to provide both calibrations.


\begin{sidewaystable*}
\caption{Optical and infrared magnitudes, colors and bolometric corrections of O stars derived using the ``observational'' \teff\ - scale of MSH05.}
\label{tab_obs}
\center
\begin{tabular}{l|ccccccccccccccccc}
\hline 
ST & $M_{U}$ & $M_{B}$ & $M_{V}$ & $M_{J}$ & $M_{H}$ & $M_{K}$ & $(U-B)_{\rm 0}$ & $(B-V)_{\rm 0}$ & $(J-H)_{\rm 0}$ & $(H-K)_{\rm 0}$ & $BC_{\rm U}$ & $BC_{\rm B}$ & $BC_{\rm V}$  & $BC_{\rm J}$ & $BC_{\rm H}$ & $BC_{\rm K}$ & \\
\hline
O3V   & -7.31 & -6.15 & -5.86 & -5.19 & -5.07 & -4.98 & -1.16 & -0.28 & -0.11 & -0.10 & -2.54 & -3.70 & -3.99 & -4.66 & -4.78 & -4.87& \\ 
O4V   & -7.01 & -5.85 & -5.57 & -4.91 & -4.79 & -4.69 & -1.15 & -0.28 & -0.11 & -0.10 & -2.42 & -3.57 & -3.85 & -4.52 & -4.63 & -4.73& \\
O5V   & -6.69 & -5.54 & -5.27 & -4.60 & -4.49 & -4.39 & -1.14 & -0.28 & -0.11 & -0.10 & -2.29 & -3.43 & -3.71 & -4.37 & -4.48 & -4.58& \\
O5.5V & -6.56 & -5.42 & -5.14 & -4.48 & -4.37 & -4.27 & -1.14 & -0.28 & -0.11 & -0.10 & -2.22 & -3.36 & -3.64 & -4.29 & -4.40 & -4.50& \\
O6V   & -6.40 & -5.27 & -4.99 & -4.34 & -4.22 & -4.13 & -1.13 & -0.28 & -0.11 & -0.10 & -2.15 & -3.28 & -3.56 & -4.21 & -4.33 & -4.42& \\
O6.5V & -6.24 & -5.12 & -4.84 & -4.19 & -4.08 & -3.98 & -1.13 & -0.27 & -0.11 & -0.10 & -2.08 & -3.21 & -3.48 & -4.13 & -4.24 & -4.34& \\
O7V   & -6.09 & -4.97 & -4.70 & -4.05 & -3.94 & -3.84 & -1.12 & -0.27 & -0.11 & -0.10 & -2.01 & -3.13 & -3.40 & -4.05 & -4.16 & -4.26& \\
O7.5V & -5.94 & -4.83 & -4.56 & -3.91 & -3.80 & -3.70 & -1.11 & -0.27 & -0.11 & -0.10 & -1.93 & -3.05 & -3.32 & -3.96 & -4.07 & -4.17& \\
O8V   & -5.79 & -4.68 & -4.41 & -3.78 & -3.67 & -3.57 & -1.11 & -0.27 & -0.11 & -0.10 & -1.86 & -2.97 & -3.24 & -3.87 & -3.98 & -4.08& \\
O8.5V & -5.62 & -4.52 & -4.25 & -3.62 & -3.51 & -3.41 & -1.10 & -0.27 & -0.11 & -0.10 & -1.78 & -2.88 & -3.15 & -3.78 & -3.89 & -3.99& \\
O9V   & -5.48 & -4.38 & -4.12 & -3.48 & -3.38 & -3.28 & -1.10 & -0.27 & -0.11 & -0.10 & -1.70 & -2.79 & -3.06 & -3.69 & -3.80 & -3.90& \\
O9.5V & -5.34 & -4.25 & -3.98 & -3.36 & -3.25 & -3.15 & -1.09 & -0.26 & -0.11 & -0.10 & -1.61 & -2.70 & -2.97 & -3.59 & -3.70 & -3.80& \\
\hline
O3III   & -7.63 & -6.47 & -6.18 & -5.51 & -5.40 & -5.30 & -1.16 & -0.28 & -0.11 & -0.10 & -2.52 & -3.68 & -3.97 & -4.64 & -4.75 & -4.85 & \\
O4III   & -7.49 & -6.33 & -6.05 & -5.39 & -5.27 & -5.18 & -1.15 & -0.28 & -0.11 & -0.10 & -2.39 & -3.54 & -3.82 & -4.49 & -4.60 & -4.70 & \\
O5III   & -7.33 & -6.18 & -5.91 & -5.25 & -5.14 & -5.04 & -1.14 & -0.28 & -0.11 & -0.10 & -2.25 & -3.39 & -3.67 & -4.33 & -4.44 & -4.54 & \\
O5.5III & -7.25 & -6.11 & -5.84 & -5.18 & -5.07 & -4.97 & -1.14 & -0.28 & -0.11 & -0.10 & -2.18 & -3.31 & -3.59 & -4.24 & -4.36 & -4.46 & \\
O6III   & -7.17 & -6.04 & -5.77 & -5.12 & -5.00 & -4.91 & -1.13 & -0.27 & -0.11 & -0.10 & -2.10 & -3.23 & -3.51 & -4.16 & -4.27 & -4.37 & \\
O6.5III & -7.07 & -5.95 & -5.68 & -5.03 & -4.92 & -4.82 & -1.12 & -0.27 & -0.11 & -0.10 & -2.03 & -3.15 & -3.42 & -4.07 & -4.18 & -4.28 & \\
O7III   & -7.00 & -5.88 & -5.61 & -4.97 & -4.86 & -4.76 & -1.12 & -0.27 & -0.11 & -0.10 & -1.95 & -3.07 & -3.34 & -3.98 & -4.09 & -4.19 & \\
O7.5III & -6.93 & -5.82 & -5.55 & -4.91 & -4.80 & -4.70 & -1.11 & -0.27 & -0.11 & -0.10 & -1.87 & -2.98 & -3.25 & -3.89 & -4.00 & -4.10 & \\
O8III   & -6.84 & -5.74 & -5.47 & -4.83 & -4.72 & -4.62 & -1.10 & -0.27 & -0.11 & -0.10 & -1.79 & -2.89 & -3.16 & -3.79 & -3.90 & -4.00 & \\
O8.5III & -6.75 & -5.65 & -5.39 & -4.76 & -4.65 & -4.55 & -1.10 & -0.27 & -0.11 & -0.10 & -1.70 & -2.80 & -3.06 & -3.69 & -3.80 & -3.90 & \\
O9III   & -6.66 & -5.58 & -5.31 & -4.68 & -4.58 & -4.48 & -1.09 & -0.26 & -0.11 & -0.10 & -1.61 & -2.70 & -2.96 & -3.59 & -3.70 & -3.80 & \\
O9.5III & -6.60 & -5.52 & -5.26 & -4.64 & -4.53 & -4.43 & -1.08 & -0.26 & -0.11 & -0.10 & -1.52 & -2.60 & -2.86 & -3.48 & -3.59 & -3.69 & \\
\hline
O3I   & -7.85 & -6.70 & -6.42 & -5.75 & -5.64 & -5.54 & -1.15 & -0.28 & -0.11 & -0.10 & -2.38 & -3.53 & -3.81 & -4.47 & -4.59 & -4.69 & \\
O4I   & -7.82 & -6.68 & -6.40 & -5.74 & -5.63 & -5.53 & -1.14 & -0.28 & -0.11 & -0.10 & -2.26 & -3.40 & -3.68 & -4.34 & -4.45 & -4.55 & \\
O5I   & -7.79 & -6.66 & -6.39 & -5.73 & -5.62 & -5.52 & -1.13 & -0.28 & -0.11 & -0.10 & -2.13 & -3.26 & -3.54 & -4.19 & -4.30 & -4.40 & \\
O5.5I & -7.78 & -6.66 & -6.38 & -5.73 & -5.62 & -5.52 & -1.13 & -0.27 & -0.11 & -0.10 & -2.07 & -3.19 & -3.47 & -4.12 & -4.23 & -4.33 & \\
O6I   & -7.77 & -6.65 & -6.38 & -5.73 & -5.62 & -5.52 & -1.12 & -0.27 & -0.11 & -0.10 & -2.00 & -3.12 & -3.40 & -4.04 & -4.15 & -4.25 & \\
O6.5I & -7.76 & -6.65 & -6.38 & -5.74 & -5.62 & -5.53 & -1.11 & -0.27 & -0.11 & -0.10 & -1.94 & -3.05 & -3.32 & -3.96 & -4.08 & -4.17 & \\
O7I   & -7.76 & -6.65 & -6.38 & -5.74 & -5.63 & -5.53 & -1.11 & -0.27 & -0.11 & -0.10 & -1.87 & -2.98 & -3.24 & -3.89 & -4.00 & -4.09 & \\
O7.5I & -7.75 & -6.65 & -6.38 & -5.75 & -5.64 & -5.54 & -1.10 & -0.27 & -0.11 & -0.10 & -1.80 & -2.90 & -3.17 & -3.80 & -3.91 & -4.01 & \\
O8I   & -7.73 & -6.63 & -6.36 & -5.73 & -5.62 & -5.52 & -1.10 & -0.27 & -0.11 & -0.10 & -1.72 & -2.82 & -3.09 & -3.72 & -3.83 & -3.93 & \\
O8.5I & -7.73 & -6.64 & -6.37 & -5.74 & -5.63 & -5.53 & -1.09 & -0.26 & -0.11 & -0.10 & -1.65 & -2.74 & -3.00 & -3.63 & -3.74 & -3.84 & \\
O9I   & -7.70 & -6.62 & -6.36 & -5.73 & -5.62 & -5.52 & -1.08 & -0.26 & -0.11 & -0.10 & -1.57 & -2.66 & -2.92 & -3.54 & -3.65 & -3.75 & \\
O9.5I & -7.68 & -6.61 & -6.34 & -5.72 & -5.62 & -5.52 & -1.08 & -0.26 & -0.11 & -0.10 & -1.49 & -2.57 & -2.83 & -3.45 & -3.56 & -3.66 & \\
\hline 
\end{tabular}
\end{sidewaystable*}


\begin{sidewaystable*}
\caption{Optical and infrared magnitudes, colors and bolometric corrections of O stars derived using the ``theoretical'' \teff\ - scale of MSH05.}
\label{tab_theo}
\center
\begin{tabular}{l|ccccccccccccccccc}
\hline 
ST & $M_{U}$ & $M_{B}$ & $M_{V}$ & $M_{J}$ & $M_{H}$ & $M_{K}$ & $(U-B)_{\rm 0}$ & $(B-V)_{\rm 0}$ & $(J-H)_{\rm 0}$ & $(H-K)_{\rm 0}$ & $BC_{\rm U}$ & $BC_{\rm B}$ & $BC_{\rm V}$  & $BC_{\rm J}$ & $BC_{\rm H}$ & $BC_{\rm K}$ & \\
\hline
O3V   & -7.30 & -6.14 & -5.85 & -5.18 & -5.07 & -4.97 & -1.16 & -0.28 & -0.11 & -0.10 & -2.52 & -3.69 & -3.97 & -4.65 & -4.76 & -4.86 & \\
O4V   & -7.00 & -5.84 & -5.56 & -4.89 & -4.78 & -4.68 & -1.16 & -0.28 & -0.11 & -0.10 & -2.45 & -3.61 & -3.89 & -4.56 & -4.67 & -4.77 & \\
O5V   & -6.69 & -5.55 & -5.27 & -4.60 & -4.49 & -4.39 & -1.15 & -0.28 & -0.11 & -0.10 & -2.33 & -3.48 & -3.76 & -4.42 & -4.53 & -4.63 & \\
O5.5V & -6.54 & -5.40 & -5.12 & -4.47 & -4.35 & -4.26 & -1.14 & -0.28 & -0.11 & -0.10 & -2.23 & -3.37 & -3.65 & -4.31 & -4.42 & -4.52 & \\
O6V   & -6.40 & -5.27 & -5.00 & -4.34 & -4.23 & -4.13 & -1.13 & -0.27 & -0.11 & -0.10 & -2.10 & -3.23 & -3.50 & -4.16 & -4.27 & -4.37 & \\
O6.5V & -6.25 & -5.12 & -4.85 & -4.20 & -4.09 & -3.99 & -1.12 & -0.27 & -0.11 & -0.10 & -2.00 & -3.13 & -3.40 & -4.05 & -4.16 & -4.26 & \\
O7V   & -6.09 & -4.98 & -4.71 & -4.07 & -3.96 & -3.86 & -1.11 & -0.27 & -0.11 & -0.10 & -1.91 & -3.02 & -3.29 & -3.93 & -4.04 & -4.14 & \\
O7.5V & -5.93 & -4.82 & -4.55 & -3.92 & -3.81 & -3.71 & -1.11 & -0.27 & -0.11 & -0.10 & -1.82 & -2.93 & -3.20 & -3.83 & -3.94 & -4.04 & \\
O8V   & -5.76 & -4.66 & -4.40 & -3.76 & -3.65 & -3.55 & -1.10 & -0.27 & -0.11 & -0.10 & -1.74 & -2.84 & -3.10 & -3.74 & -3.85 & -3.95 & \\
O8.5V & -5.63 & -4.54 & -4.27 & -3.64 & -3.53 & -3.44 & -1.09 & -0.26 & -0.11 & -0.10 & -1.67 & -2.76 & -3.03 & -3.66 & -3.77 & -3.86 & \\
O9V   & -5.47 & -4.38 & -4.12 & -3.49 & -3.38 & -3.28 & -1.09 & -0.26 & -0.11 & -0.10 & -1.58 & -2.67 & -2.93 & -3.56 & -3.67 & -3.77 & \\
O9.5V & -5.31 & -4.23 & -3.97 & -3.35 & -3.24 & -3.14 & -1.08 & -0.26 & -0.11 & -0.10 & -1.49 & -2.57 & -2.83 & -3.45 & -3.56 & -3.66 & \\
\hline
O3III   & -7.63 & -6.47 & -6.19 & -5.52 & -5.41 & -5.31 & -1.16 & -0.28 & -0.11 & -0.10 & -2.42 & -3.58 & -3.86 & -4.53 & -4.64 & -4.74 & \\
O4III   & -7.47 & -6.33 & -6.05 & -5.38 & -5.27 & -5.17 & -1.15 & -0.28 & -0.11 & -0.10 & -2.33 & -3.47 & -3.75 & -4.42 & -4.53 & -4.63	& \\
O5III   & -7.30 & -6.17 & -5.89 & -5.24 & -5.12 & -5.02 & -1.14 & -0.28 & -0.11 & -0.10 & -2.20 & -3.33 & -3.61 & -4.26 & -4.38 & -4.48	& \\
O5.5III & -7.23 & -6.11 & -5.83 & -5.18 & -5.07 & -4.97 & -1.13 & -0.27 & -0.11 & -0.10 & -2.09 & -3.22 & -3.49 & -4.14 & -4.25 & -4.35	& \\
O6III   & -7.16 & -6.04 & -5.76 & -5.12 & -5.01 & -4.91 & -1.12 & -0.27 & -0.11 & -0.10 & -1.99 & -3.11 & -3.39 & -4.03 & -4.14 & -4.24	& \\
O6.5III & -7.06 & -5.95 & -5.67 & -5.03 & -4.92 & -4.82 & -1.11 & -0.27 & -0.11 & -0.10 & -1.92 & -3.03 & -3.30 & -3.94 & -4.05 & -4.15	& \\
O7III   & -6.99 & -5.88 & -5.61 & -4.97 & -4.86 & -4.76 & -1.11 & -0.27 & -0.11 & -0.10 & -1.84 & -2.95 & -3.21 & -3.85 & -3.96 & -4.06	& \\
O7.5III & -6.90 & -5.80 & -5.54 & -4.90 & -4.79 & -4.69 & -1.10 & -0.27 & -0.11 & -0.10 & -1.75 & -2.85 & -3.11 & -3.75 & -3.86 & -3.96	& \\
O8III   & -6.83 & -5.73 & -5.47 & -4.84 & -4.73 & -4.63 & -1.09 & -0.26 & -0.11 & -0.10 & -1.67 & -2.77 & -3.03 & -3.66 & -3.77 & -3.87	& \\
O8.5III & -6.75 & -5.66 & -5.40 & -4.77 & -4.67 & -4.57 & -1.09 & -0.26 & -0.11 & -0.10 & -1.60 & -2.69 & -2.95 & -3.58 & -3.68 & -3.78	& \\
O9III   & -6.66 & -5.58 & -5.32 & -4.70 & -4.59 & -4.49 & -1.08 & -0.26 & -0.11 & -0.10 & -1.52 & -2.60 & -2.86 & -3.48 & -3.59 & -3.69	& \\
O9.5III & -6.58 & -5.50 & -5.24 & -4.62 & -4.51 & -4.42 & -1.08 & -0.26 & -0.11 & -0.10 & -1.47 & -2.55 & -2.81 & -3.43 & -3.54 & -3.63	& \\
\hline
O3I   & -7.85 & -6.70 & -6.42 & -5.75 & -5.64 & -5.54 & -1.15 & -0.28 & -0.11 & -0.10 & -2.40 & -3.55 & -3.83 & -4.50 & -4.61 & -4.71 & \\
O4I   & -7.82 & -6.68 & -6.40 & -5.74 & -5.63 & -5.53 & -1.14 & -0.28 & -0.11 & -0.10 & -2.28 & -3.42 & -3.70 & -4.36 & -4.47 & -4.57 & \\
O5I   & -7.80 & -6.67 & -6.39 & -5.74 & -5.63 & -5.53 & -1.13 & -0.28 & -0.11 & -0.10 & -2.13 & -3.26 & -3.53 & -4.19 & -4.30 & -4.40 & \\
O5.5I & -7.78 & -6.66 & -6.38 & -5.73 & -5.62 & -5.52 & -1.12 & -0.27 & -0.11 & -0.10 & -2.02 & -3.14 & -3.42 & -4.07 & -4.18 & -4.28 & \\
O6I   & -7.78 & -6.66 & -6.39 & -5.75 & -5.64 & -5.54 & -1.11 & -0.27 & -0.11 & -0.10 & -1.92 & -3.04 & -3.31 & -3.95 & -4.06 & -4.16 & \\
O6.5I & -7.76 & -6.65 & -6.38 & -5.75 & -5.64 & -5.54 & -1.11 & -0.27 & -0.11 & -0.10 & -1.84 & -2.95 & -3.22 & -3.85 & -3.96 & -4.06 & \\
O7I   & -7.74 & -6.64 & -6.38 & -5.74 & -5.63 & -5.53 & -1.10 & -0.27 & -0.11 & -0.10 & -1.73 & -2.83 & -3.10 & -3.73 & -3.84 & -3.94 & \\
O7.5I & -7.73 & -6.64 & -6.38 & -5.75 & -5.64 & -5.55 & -1.09 & -0.26 & -0.11 & -0.10 & -1.62 & -2.71 & -2.97 & -3.60 & -3.71 & -3.80 & \\
O8I   & -7.71 & -6.63 & -6.37 & -5.74 & -5.63 & -5.54 & -1.08 & -0.26 & -0.11 & -0.10 & -1.54 & -2.62 & -2.88 & -3.51 & -3.62 & -3.71 & \\
O8.5I & -7.70 & -6.63 & -6.37 & -5.74 & -5.64 & -5.54 & -1.08 & -0.26 & -0.11 & -0.10 & -1.50 & -2.57 & -2.83 & -3.46 & -3.56 & -3.66 & \\
O9I   & -7.69 & -6.62 & -6.36 & -5.74 & -5.63 & -5.54 & -1.07 & -0.26 & -0.11 & -0.10 & -1.41 & -2.48 & -2.74 & -3.36 & -3.47 & -3.56 & \\
O9.5I & -7.67 & -6.61 & -6.35 & -5.74 & -5.63 & -5.53 & -1.06 & -0.26 & -0.11 & -0.10 & -1.30 & -2.37 & -2.62 & -3.24 & -3.34 & -3.44 & \\
\end{tabular}
\end{sidewaystable*}

\subsection{Accuracy of calibrations}
\label{uncertainties_cal}

\begin{figure}
\centerline{\psfig{file=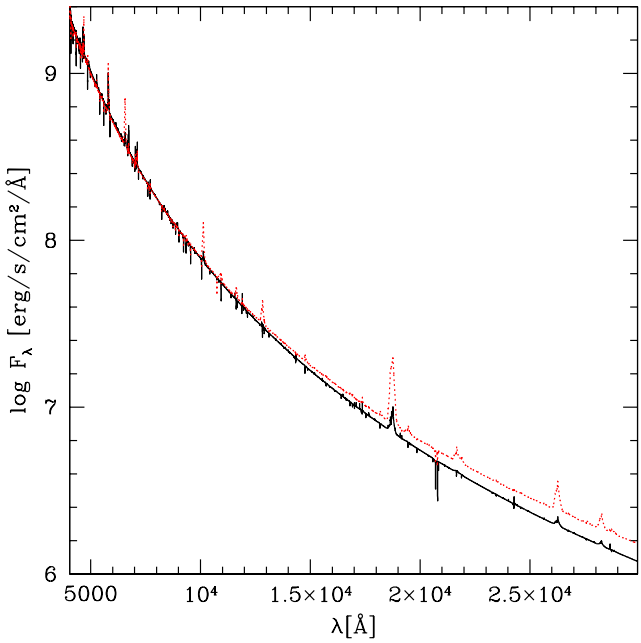,width=9cm}}
\caption{Variation of the SED as a function of mass loss rate: the
black solid curve is for \mdot=$1.32 \times 10^{-5}$ \myr\ while the red dotted
line is for a mass loss rate three times larger. The other parameters
of the model are \teff=48530 K, \logg=4.01 and \vinf=2977 \kms.  }
\label{fig_mdotX3}
\end{figure}

In Fig.\ \ref{BC_teff} and Table \ref{bc_teff_param}, we see that the
dispersions around the average $BC - \teff$ relation increases with
wavelength. This is a natural consequence of the stronger sensitivity
of the SED to wind parameters at longer wavelengths. Indeed, massive
stars are known to emit a significant excess of radiation in the
IR-radio range due to free-free emission originating in their wind
\citep[e.g.][]{lcbook}. This is illustrated in Fig.\ \ref{fig_mdotX3}
which shows the variation of the SED when the mass loss rate is
increased by a factor of 3 (which is the typical uncertainty claimed
by detailed analysis) in a model for an early supergiant. This reveals
that any variation in the wind parameters will change the wind
density, which in turn will affect both the continuum level and the
strength of near IR emission lines (see Fig.\ \ref{fig_mdotX3}),
leading to a modification of the IR photometry. In order to estimate
the magnitude of this effect, we have run test models for a sample of
stars for which the mass loss rate was increased by a factor of 3. The
results are gathered in Table \ref{tab_mdotX3}.  For such a change in
wind density the typical variation in bolometric correction is $\sim$
0.02 mag in U, while it can reach 0.2 mag in K. Note that this is
similar to the dispersion of relation \ref{eq_reg_bc}.

\begin{table}
\caption{Bolometric corrections of models with \mdot\ increased by a
factor of 3 (A2,B2,C2,D2) compared to initial models
(A1,B1,C1,D1). The parameters of the initial models are: \teff=32210
K, \logg=3.26, \mdot=$10^{-4.93}$ \myr, \vinf=1960 \kms\ for model A1,
\teff=33340 K, \logg=4.01, \mdot=$10^{-6.86}$ \myr, \vinf=2544 \kms\
for model B1, \teff=42560 K, \logg=3.71, \mdot=$10^{-4.88}$ \myr,
\vinf=2538 \kms\ for model C1, \teff=48530 K, \logg=4.01,
\mdot=$10^{-5.38}$ \myr, \vinf=2977 \kms\ for model D1.}
\label{tab_mdotX3}
\center
\begin{tabular}{l|lllllll}
\hline 
    & $BC_{\rm U}$ & $BC_{\rm B}$ & $BC_{\rm V}$  & $BC_{\rm J}$ & $BC_{\rm H}$ & $BC_{\rm K}$ \\
\hline
A1 & -1.644 & -2.738 & -2.984 & -3.544 & -3.636 & -3.712 \\
A2 & -1.654 & -2.750 & -3.006 & -3.519 & -3.588 & -3.607 \\
B1 & -1.685 & -2.799 & -3.085 & -3.773 & -3.893 & -4.003 \\
B2 & -1.685 & -2.800 & -3.087 & -3.776 & -3.892 & -3.999 \\
C1 & -2.442 & -3.592 & -3.865 & -4.506 & -4.603 & -4.681 \\
C2 & -2.460 & -3.609 & -3.881 & -4.441 & -4.500 & -4.496 \\
D1 & -2.815 & -3.980 & -4.272 & -4.958 & -5.071 & -5.174 \\
D2 & -2.806 & -3.980 & -4.268 & -4.941 & -5.039 & -5.115 \\
\hline 
\end{tabular}
\end{table}

Fig.\ \ref{BC_teff} also shows the calibration of $BC_{V}$ as a
function of \teff\ obtained by MSH05. We see that it is slightly
different from the one presented here, although it is based on the
same models. The reason for this small discrepancy is the use of a
fixed effective wavelength (5500 \AA) in the computation of MSH05,
while here we use filter curves providing a better measure of the flux
in the $V$ band.
The corresponding new effective
wavelengths are usually shorter than 5500 \AA, leading to a slightly
smaller magnitude and consequently to a slightly larger bolometric
correction, as seen in Fig.\ \ref{BC_teff}. Recomputing photometry as
in MSH05 with the more accurate effective wavelengths of the present
study leads to excellent agreement with the magnitudes presented here
(differences smaller than 0.01 mag).


In conclusion, we see that the accuracy of the magnitudes and
bolometric corrections is usually better than 0.1 mag and reaches 0.2 mag
in K band. In any case, this uncertainty is of the order of the
difference between two spectral sub types within a luminosity class.

%
%
\section{Comparison with previous studies}
\label{comp_previous}

\subsection{Absolute magnitudes}
\label{disc_mag}

\begin{figure}
\centerline{\psfig{file=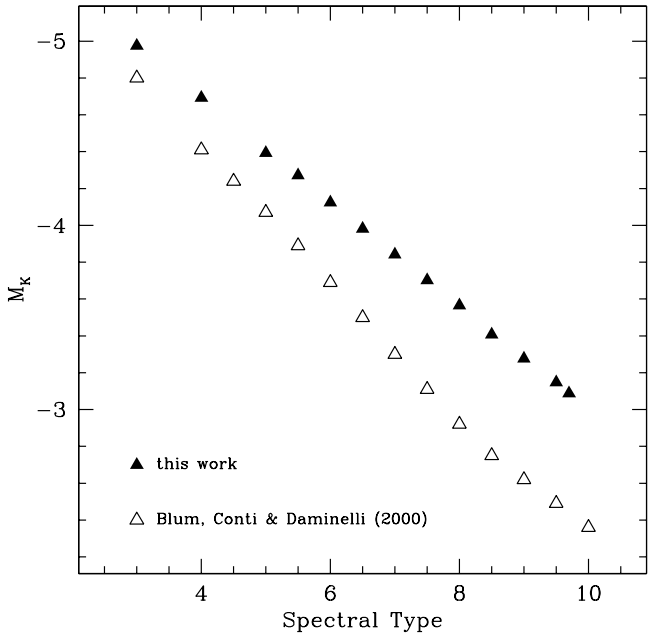,width=9cm}}
\caption{Absolute K magnitudes as a function of spectral types for
dwarfs. Filled triangles are this work, while open triangles are the
calibration of \citet{bcd00}. The difference is mostly due to the fact
that \citet{bcd00} give parameters for ZAMS stars while we provide
calibration for ``normal'' O dwarfs.}
\label{comp_MK}
\end{figure}

In terms of infrared magnitudes, \citet{bcd00} derived absolute K
magnitudes for ZAMS O stars using evolutionary models
\citep{schaller92}, $V-K$ from \citet{kor83}, and $BC_{V} - ST$ and
\teff\ $- ST$ relations of \citet{vacca}. The comparison between their
results and our relation for dwarfs is displayed in Fig.\
\ref{comp_MK}. We see that our calibration is slightly brighter. Since
we calibrated photometry for ``normal'' O stars, it is not surprising
that the ZAMS K magnitudes given by \citet{bcd00} are fainter, ZAMS
stars being less bright than standard O dwarfs
\citep{hanson97,niemela06}. Note that the difference between the two
calibrations increases with spectral type, which is also what is seen
when comparing the position of standard O stars to the ZAMS in a HR
diagram (e.g. Fig.\ 12 in MSH05). One could also argue that the use of
different \teff\ - scales is responsible for the observed
differences. However, this is not the case: had we used the same
approach as \citet{bcd00} to derive K magnitudes of ZAMS stars, but
using the recent \teff\ scale of MSH05, we would have found
\textit{fainter} K magnitudes than they find. Indeed, for a given ZAMS
evolutionary model, i.e. for a given \teff\ and luminosity, $BC_{V}$
is similar if we use the \citet{vacca} $BC_{V}$ - \teff\ calibration
or the present one (similar to MSH05 too): the difference in
bolometric corrections is less than 0.1 mag. Hence we would derive a
similar V magnitude (within 0.1 mag) for the model. Using $V-K$ from
\citet{kor83} to get the K magnitude would then lead to a similar
$M_{\rm K}$ (again within 0.1 mag). But now, using the recent \teff\ -
scale of MSH05 instead of the one by \citet{vacca} \citep[as
in][]{bcd00} to convert \teff\ into spectral type gives earlier
values: the effective temperatures of MSH05 are indeed cooler for a
given spectral type. This means that using the effective temperature
calibrations of MSH05 instead of that of \citet{vacca} translates to a
shift of the $M_{\rm K} - ST$ relation towards earlier spectral
types. In that case, the difference with the calibration we provide in
the present work is even higher. Hence, the use of different \teff\ -
scales is not responsible for the shift seen in Fig.\ \ref{comp_MK}
which is more likely attributed to the different evolutionary status
in the stars considered (ZAMS O stars versus ``normal'' O stars).

Calibrations of the $V$ magnitudes as a function of spectral type have
been discussed in MSH05 (see their Sect. 5.1): reasonable agreement
with previous calibrations (differences smaller than 0.4 mag) was
found. Inspection of Tables \ref{tab_obs} and \ref{tab_theo} together
with Tables 1-6 of MSH05 reveals that the present V magnitudes are
systematically smaller by 0.02-0.08 magnitudes which is simply due to
the better computation of the $V$ band photometry in the present study
as discussed in Sect.\ \ref{uncertainties_cal}. This difference is
however well within the typical uncertainty of the calibrations.

\subsection{Intrinsic colors}
\label{disc_colors}

\subsubsection{Optical range}
\label{opt_col}

\begin{figure}
\centerline{\psfig{file=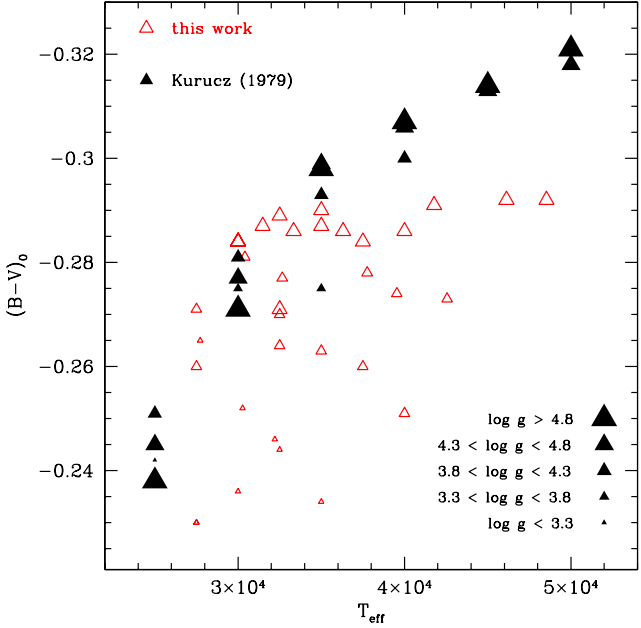,width=9cm}}
\caption{$(B-V)_{0}$ as a function of effective temperature
for the present work (red open triangles) and from \citet{kur79}
(black full triangles). For the present work, $(B-V)_{0}$ have been
computed from the original grid of models and are not taken
directly from Tables \ref{tab_obs} and \ref{tab_theo}. The size of the
symbols scales with \logg\ (bigger symbols corresponding to larger
gravities).}
\label{comp_BmV}
\end{figure}

Fig.\ \ref{comp_BmV} shows the relation $(B-V)_{0} - \teff$ from two
sets of atmosphere models: the CMFGEN \footnote{For more details on
the models, see \citet{hm98}.} models of MSH05 and Kurucz models
\citep{kur79}. On average and for a given \teff, it seems that the new
colors are slightly redder, and that the dispersion is larger
too. However, this can be explained simply by the effect of
gravity. As shown by \citet{ah85} (and confirmed here), the SED of O
stars depends on \logg: when gravity decreases, the optical flux
distributions get redder. This is clearly shown in Fig.\
\ref{comp_BmV} where for a given \teff\ a clear sequence of bluer
$(B-V)_{0}$ appears when \logg\ is increased \footnote{Note however
that for the models at low \teff, deviations to this general trend
appear.}. The range of \logg\ covered by the Kurucz models is
3.50-5.00, while the grid of MSH05 has 3.20 $<$ \logg\ $<$ 4.25 (see
coding of symbol size in Fig.\ \ref{comp_BmV}). Thus on average,
redder $(B-V)_{0}$ are naturally expected for these new models (as
seen in Fig.\ \ref{comp_BmV}). The models of MSH05 also do not reach
very blue colors since they are restricted to lower gravities compared
to the Kurucz models. It is also worth noting that the CMFGEN models
include winds, which are know to affect the SED \citep[see Sect.\
\ref{uncertainties_cal} and][]{gabler89}. However, the effects are
reduced in the optical range. A simple inspection of Table
\ref{tab_mdotX3} reveals that increasing the mass loss rate by a
factor 3 leads to changes in $B-V$ (equal to $BC_{V} - BC_{B}$ for a
given model) by no more than 0.1 mag. Consequently, the redder colors
observed in Fig.\ \ref{comp_BmV} for the new models can safely be
attributed to lower \logg.

Further inspection of Tables \ref{tab_obs} and \ref{tab_theo}
reveals that for standard O stars of any spectral type / luminosity
class, \textit{$(B-V)_{0}$ is never bluer than -0.28}. This is because
normal O stars have well defined $\logg\ - ST$ relations for which
\logg\ never exceeds $\sim$ 4.0 (see e.g. Fig.\ 2 in MSH05). Again,
this does not mean that very blue (lower than -0.30) $(B-V)_{0}$
colors can not be obtained with atmosphere models, but this requires
very large gravities (\logg\ $>$ 4.25) which are not typical of normal
O stars (but probably more of very young massive stars).

How does this result compare to previous studies?
\citet{johnson66} derived $(B-V)_{0}=-0.30/-0.32$ for O5-O9.5 stars
fitting the envelope of the position of O stars in a two colors
diagram (see his Fig. 1). This method assumes that the color is almost
independent of spectral type (which is indeed the case) but Johnson
uses a sample of stars covering only part of the range of spectral types
(especially missing the earliest O stars) which may lead to
uncertainties in the derived average values.

\citet{fitz70} determined intrinsic colors of O stars also using the method
of the blue-most envelope in two colors diagram, but he adopted
$(B-V)_{0}=-0.32$ from \citet{johnson63}. This color was determined
from studies of O stars in associations in which the color excess was
determined from photometry of later type stars.

Later, \citet{cgm86} derived colors of early type stars in the LMC and
found that O stars had $(B-V)_{0}=-0.30$ for most O stars and
$(B-V)_{0}=-0.24$ for late supergiants. They also highlighted that the
intrinsic colors in the LMC were slightly redder than in the Galaxy
(while they should be similar or slightly bluer due to
metallicity effects, see below), pointing to a possible problem with
the Galactic reddening estimates, and consequently with the Galactic
intrinsic colors.

\citet{fitz88} also studied OB supergiants in the LMC and derived
$(B-V)_{0}=-0.27$ for O3-6 stars and $(B-V)_{0}=-0.23$ for O7-9.7
supergiants, stressing the variation of colors within the different O
stars luminosity classes. \citet{fitz88} also compared Str$\ddot{\rm
o}$mgren and Johnson intrinsic colors in the Galaxy and LMC and
highlighted again possible calibration uncertainties for the Galaxy.

We find $(B-V)_{0} > -0.28$ for Galactic O stars. This is about 0.04
mag redder than the early results of \citet{johnson66} and
\citet{fitz70}, but in reasonable agreement with the more recent study
of \citet{fitz88}, although this study is based on LMC stars. We have
not computed models at $Z = 0.5 Z_{\odot}$ (typical of the LMC), but
we do not expect changes in $(B-V)_{0}$ larger than $\sim$ 0.01
mag. Indeed, effective temperatures are expected to be larger - for a
given spectral type - in low metallicity environments \citep[see
e.g.][]{mokiem04,massey04}, but examination of Tables \ref{tab_obs}
and \ref{tab_theo} shows that $(B-V)_{0}$ is very little sensitive to
\teff\ within the whole range of O stars: it changes by no more than
0.02 mag. Hence, $(B-V)_{0}$ should be very similar in the LMC and in
the Galaxy which is confirmed by the good agreement between our
results and those of \citet{fitz88}. Note however that we find
$(B-V)_{0}$ slightly bluer than \citet{fitz88} for late
supergiants. The origin of this difference is not clear at
present. ``Wind effects'' can be excluded since weaker winds -
expected at lower metallicity - should lead to bluer SEDs and colors
(although the effect is small in the optical, see above discussion),
which is the opposite of what we see.

Such a change of the intrinsic $B-V$ colors of standard O stars
\footnote{We stress again that $(B-V)_{0}$ bluer than -0.28 can be
obtained in atmosphere models if large gravities (\logg\ $\gtrsim$ 4.25)
are encountered, but such gravities are not typical of normal O stars}
(at least -0.28 instead of -0.32) has important consequences for the
distance determination of OB associations and young clusters.
An increase of 0.04 mag in $(B-V)_{0}$ translates to a decrease of
$E(B-V)$ by the same amount, which then implies a reduction of $A_{V}$
by $\sim$ 0.124 mag (adopting $R_{V}$ = 3.1). This means that the
distance modulus is reduced by the same amount. For clusters such as
Tr16 in the Carina region \citep[e.g.][]{degioia}, this is equivalent
to a reduction of the distance by 5.5 $\%$. Inversely, if the distance
is known, a reduction of $E(B-V)$ by 0.04 mag implies a luminosity
lower by 0.05 dex for a given star.\\

\begin{figure}
\centerline{\psfig{file=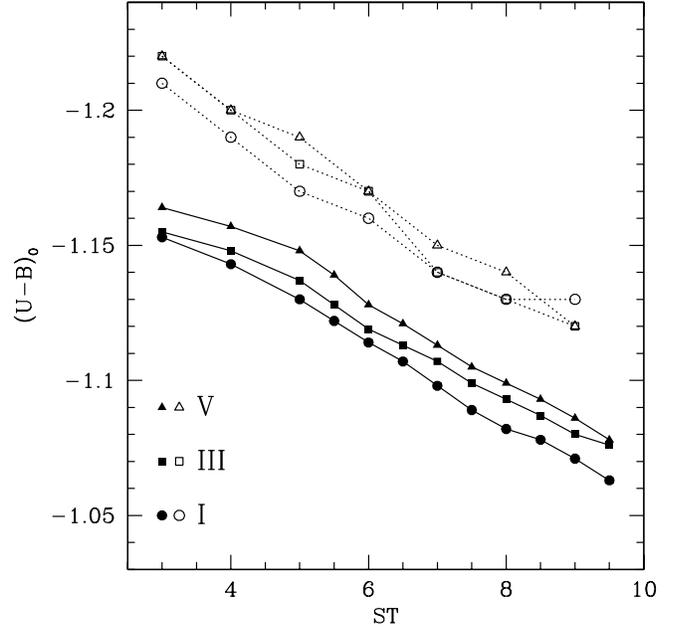,width=9cm}}
\caption{(U-B) intrinsic colors as a function of spectral type for
the present work (solid symbols) and from \citet{sk82}. Triangles
(squares, circles) are for dwarfs (giants, supergiants). See text for
discussion.}
\label{comp_UmB}
\end{figure}

Fig.\ \ref{comp_UmB} shows the comparison between $(U-B)_{0}$ colors
derived in the present study as a function of spectral type together
with the values of \citet{sk82}. Our values are systematically redder
by $\sim$ 0.05 mag. \citet{fitz70} derived $(U-B)_{0}$ between -1.19
and -1.10 for dwarfs (smaller values for early spectral types) and
between -1.13 and -1.07 for O9-9.7 giants and supergiants, in marginal
agreement with our values. \citet{cgm86} found similar values (-1.12
to -1.07) for LMC stars and give $(U-B)_{0}$=-1.17 for Galactic stars,
arguing again that this value may be too blue (see discussion
above). Finally, \citet{fitz88} derived $(U-B)_{0}=-1.04$ (resp.
$-1.08$) for O3-6 (resp. O7-9.7) supergiants.  \citet{bcp98} derived
$(U-B)_{0}$ colors for O dwarfs using the ATLAS9 atmosphere models of
\citet{kur93,kur94}. Their values range between -1.22 for the earliest
O stars to -1.07 for the latest ones (see their Table 9). This is
bluer than our results and similar to the results of \citet{sk82}, but
can be explained by the different \teff\ - scales used: adopting the
cooler $\teff\ - ST$ relation of MSH05 compared to the relation of
\citet{paul98} used by \citet{bcp98} translates into a shift of the
$(U-B)_{0} - ST$ relation towards earlier spectral types. Such a shift
(of about one spectral sub type) reduces - although not completely -
the difference between our relation and those of \citet{bcp98} and
\citet{sk82}, as seen in Fig.\ \ref{comp_UmB}. The origin of the
remaining difference is attributed to the use of more realistic models
in our study.

On average, our $(U-B)_{0}$ colors are thus consistent with previous
studies given the improvement in the model atmospheres.

\subsubsection{Near-IR range}
\label{nir_col}

There are less studies of near infrared colors of O stars than there
are for optical colors. The main ones are by \citet{johnson66},
\citet{wvb80} and \citet{kor83}. As for optical colors,
\citet{johnson66} used the method of the two colors diagram to derive
intrinsic $(J-K)_{0}$. \citet{wvb80} proceeded differently and
determined intrinsic near IR colors from a sample of 65 stars with
low extinction (as deduced from optical colors). Their work was done
in the Glass system \citep{glass74}. Finally, \citet{kor83} used a
compilation of 127 standard stars in the Southern hemisphere to
produce near-IR photometry and colors in the Johnson-Glass system (J
and K in the Johnson system, H in the Glass system).

\begin{figure}
\centerline{\psfig{file=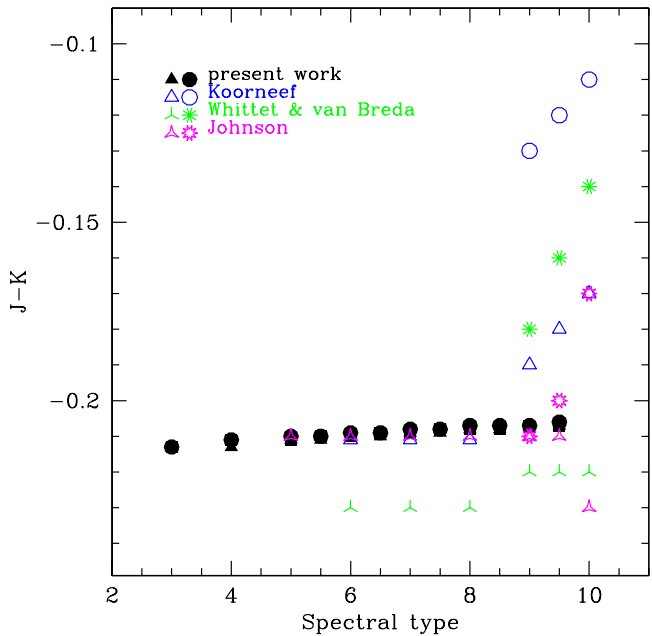,width=9cm}}
\caption{$(J-K)_{0}$ as a function of spectral type for dwarfs
(triangular shapes) and supergiants (circular shapes) from
the present work, \citet{kor83}, \citet{wvb80} and \citet{johnson66}.}
\label{comp_JmK}
\end{figure}

\begin{figure}
\centerline{\psfig{file=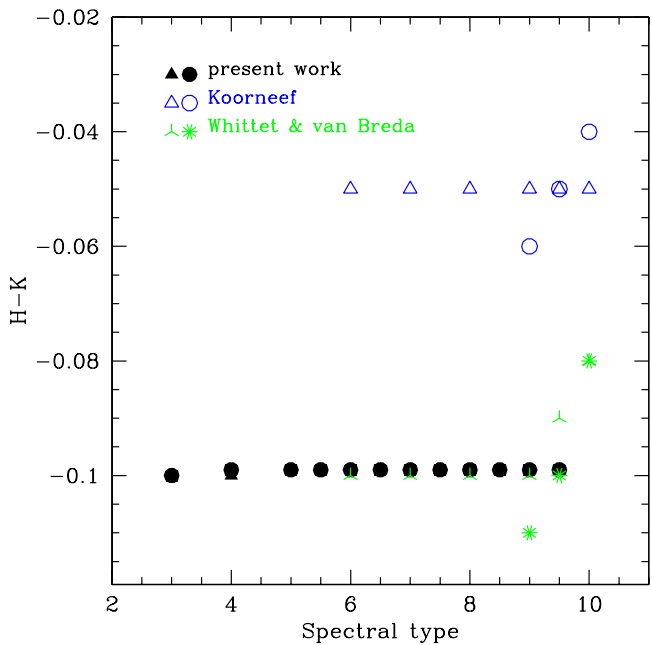,width=9cm}}
\caption{$(H-K)_{0}$ as a function of spectral for dwarfs
(triangle shapes) and supergiants (circle shapes) from
the present work, \citet{kor83} and \citet{wvb80}.}
\label{comp_HmK}
\end{figure}

Fig.\ \ref{comp_JmK} and \ref{comp_HmK} show the infrared $(J-K)_{0}$
and $(H-K)_{0}$ colors as a function of spectral type for the present
work and previous studies by \citet{johnson66}, \citet{wvb80} and
\citet{kor83}. The first interesting point to mention is that
\textit{our study covers the whole range of spectral types and
luminosity classes} which was not the case previously. Second, our IR
colors are almost constant through the range of O stars: we do not
find any significant variation of near-IR colors with spectral type or
luminosity class. 

The values of $(J-K)_{0}$ we find are similar to the
ones by \citet{johnson66}. This is also true for the colors of
\citet{kor83}, except for late spectral types for which our colors are
bluer (by $\lesssim$ 0.1 mag), especially for supergiants. The values
of \citet{wvb80} are a little bluer than ours for dwarfs, and redder
for supergiants.

As for $(H-K)_{0}$, there is a reasonably good agreement
with the values of \citet{wvb80}, whereas \citet{kor83} predicts
colors redder than ours. However, the difference is rather small
($\sim$ 0.05mag). \citet{bcd00} studied the stellar content of the HII
region W42 and give intrinsic $H-K$ colors for ZAMS stars (see their
Table 1). However, for O stars they adopt the value of \citet{kor83}.

The little differences we observe are likely partly
attributed to different photometric systems. However, the main conclusion is that
contrary to previous studies, we do not find any difference in near-IR
colors between dwarfs and supergiants.

Overall, our results are an improvement over previous theoretical
analysis due to the inclusion of line-blanketing and winds in the
atmosphere models, as well as the use of a better \teff\ - scale. They
present the advantage of covering the whole range of spectral types
and luminosity classes of O stars in a consistent way. They also
provide for the first time calibrations of bolometric corrections as a
function of spectral type for the near infrared range.

%
%
\section{Conclusion}
\label{conclusion}

We have derived calibrations of UBVJHK photometry of O stars as a
function of effective temperature and spectral type using the recent
grid of atmosphere models of \citet{msh05}. UBVJHK photometry was
computed as in \citet{bcp98}, using the system of \citet{bb88} (near
IR) and \citet{bessel90} (optical).

We provide the first calibrations of near-IR photometry, including
bolometric corrections, covering the whole range of spectral types and
luminosity classes of O stars. Infrared colors are almost
constant. $(H-K)_{0}$ is found to be -0.10, slightly lower (0.05 mag)
than the value of \citet{kor83}.

Optical photometry is consistent with recent studies. One exception is
the minimum value of $(B-V)_{0}$ for standard O stars (i.e. with \logg
$\lesssim$ 4.0) which is found to be -0.28, slightly larger than
previously accepted (-0.32). This is important when estimating
reddening and distances of OB associations since an error of 0.04 mag
in color excess amounts to an error of $\sim$ 0.1 mag in distance
modulus (or $\sim 0.05$ dex in luminosity).

These calibrations will be useful to study young massive stars
embedded in star forming regions and to better understand their
formation process.

\begin{acknowledgements}
 
FM acknowledges support from the Alexander von Humboldt foundation. We
thank an anonymous referee for his/her rapid answer.

\end{acknowledgements}

\end{document}